\documentclass[12pt,epj,twocolumn]{webofc}
\usepackage[varg]{txfonts}   

\usepackage{graphicx}
\usepackage{subfigure}
\usepackage{geometry}

\wocname{ISVHECRI 2014}
\woctitle{Constrains of hadronic interaction models from the cosmic muon observations.}
\begin{document}
\title{Constrains of hadronic interaction models from the cosmic muon observations.}
\author{Dedenko L.G.\inst{1,2}\fnsep\thanks{\email{ddn@dec1.sinp.msu.ru}} \and
        \textbf{Lukyashin A.V.}\inst{1}\fnsep\thanks{\email{lukyashin.anton@physics.msu.ru}} \and
        Fedorova G.F.\inst{2}\fnsep\thanks{\email{fdr@dec1.sinp.msu.ru}} \and
        Roganova T.M.\inst{2}\fnsep\thanks{\email{rogatm@yandex.ru}} 
}
\institute{Faculty of Physics M.V. Lomonosov Moscow State University, Leninskie Gory, 119991 Moscow, Russia.
\and Skobeltsin Institute of Nuclear Physics, Lomonosov Moscow State University, 119234 Moscow, Russia. }

\abstract{
A simple method of the vertical muon energy spectrum simulations have been suggested. 
  These calculations have been carried out in terms of various models of hadronic interactions. 
  The most energetic $ \pi^\pm $-mesons and K$^\pm $-mesons produced in hadron interactions contribute mainly in to this energy spectrum of muons due to the very steep energy spectrum of the primary particles. 
  So, some constraints on the hadronic interaction models may be set from a comparison of calculated results with the cosmic data on the vertical muon energy spectrum. 
  This comparison showed that the most energetic secondary particles production is too high in case of the QGSJET~II-04 model and rather low in case of the QGSJET~II-03 model. 
  These conclusion have been supported by the LHC data.
}
\maketitle

\section{Introduction}
\label{intro}
The longitudinal development of extensive air showers (EAS), in particular, the depth $X_{max} $ of its maximum determined by the rate of fragmentation of energy $E_{0}$  of the primary particle. 
This rate depends on the interaction cross sections of shower particles, and on the energy spectra of secondary particles are generated in interactions. 
Obviously, if the probability of  particle production with energies close to the energy of the incident particle is high then the development of the cascade slows down. 
Conversely, in the case of rapid fragmentation of the incident energy cascade develop rather rapidly. 
The depth $X_{max} $ of shower maximum  in many studies is the main parameter for determining the composition of primary cosmic radiation (PCR) at ultra-high energies. 
It should also be noted that in the case of the slow rate of development cascade and hence large values of depth $X_{max} $ lateral distribution of the shower particles at the observation level becomes narrower. 
Therefore, the values of signals in the surface and underground detectors located at large distances from the shower core are decreased. 
It must be taken into account when determining the density of muons at large distances from the shower core and the composition of the PCR found from the muon lateral distribution. 
Studies of the composition and characteristics of the energy spectrum of the PCR are important components of theories of the origin of cosmic rays at ultra-high energies. 
Interpretation of experimental data on the depth of the shower maximum $X_{max} $ and the observed fraction of muons at a fixed distance from the shower core are carried out in terms of different hadronic interaction models. 
In the case of "soft" interaction generation of secondary particles emitted at mostly small angles with respect to the projectile particle (the highest values of pseudorapidity $\eta$) is of importance. 
The models based on the Gribov-Regge theory [1, 2] are commonly used. 
The dominant contribution of the pomeron at ultra-high energies and other effects are taken into account differently in various models [3-5]. 
Therefore, testing of models at the highest energies of secondary particles is very important for understanding of physics of hadron interactions and for the interpretation of the EAS data. 
At the LHC this testing is carried out in the experiments LHCf~[6] and TOTEM~[7]. 
In cosmic rays, we suggest testing of models of hadron interactions with the help of atmospheric muon spectrum. 
In this case, due to the higher slope of the energy spectrum of the primary particles (a coefficient of the slope of the difference spectrum is $\gamma = 2.75$) generation of secondary particles ($\pi^{\pm}$-mesons and $K^{\pm}$-mesons) with the highest energies is of importance. 
It was shown [8] that the model QGSJET~II-03~[4] leads to a spectrum of vertical muons, the intensity of which is about a factor $f = 1.5$ times less than the data of collaborations L3~+~Cosmic~[9], MACRO~[10] and LVD~[11]. 
The result [8] was obtained as a solution of transport equations. 
The $\pi^{\pm}$ and $K^{\pm}$ mesons decay into $\mu^{\pm}$-mesons. 
These $\pi^{\pm}$ and $K^{\pm}$ mesons are generated by the parent particles of several generations. 
Therefore, if some excess of the secondary particle production in singular interaction of hadrons is determined by the coefficient $k$ , then, in the case of $i$ generations, the factor $f$ will be $f \approx k^i$. 
Thus, the spectra of atmospheric muons are very useful tool to test the models of hadronic interactions. 
However, other factors (cross sections of interactions, etc.) are affects result of comparison. 
In this article we propose to test the hadron interaction models by the very simple original method [12] with the help of observed atmospheric muon spectrum.
\begin{figure}
\centering
\sidecaption
\includegraphics[width=8cm,height=6cm,clip]{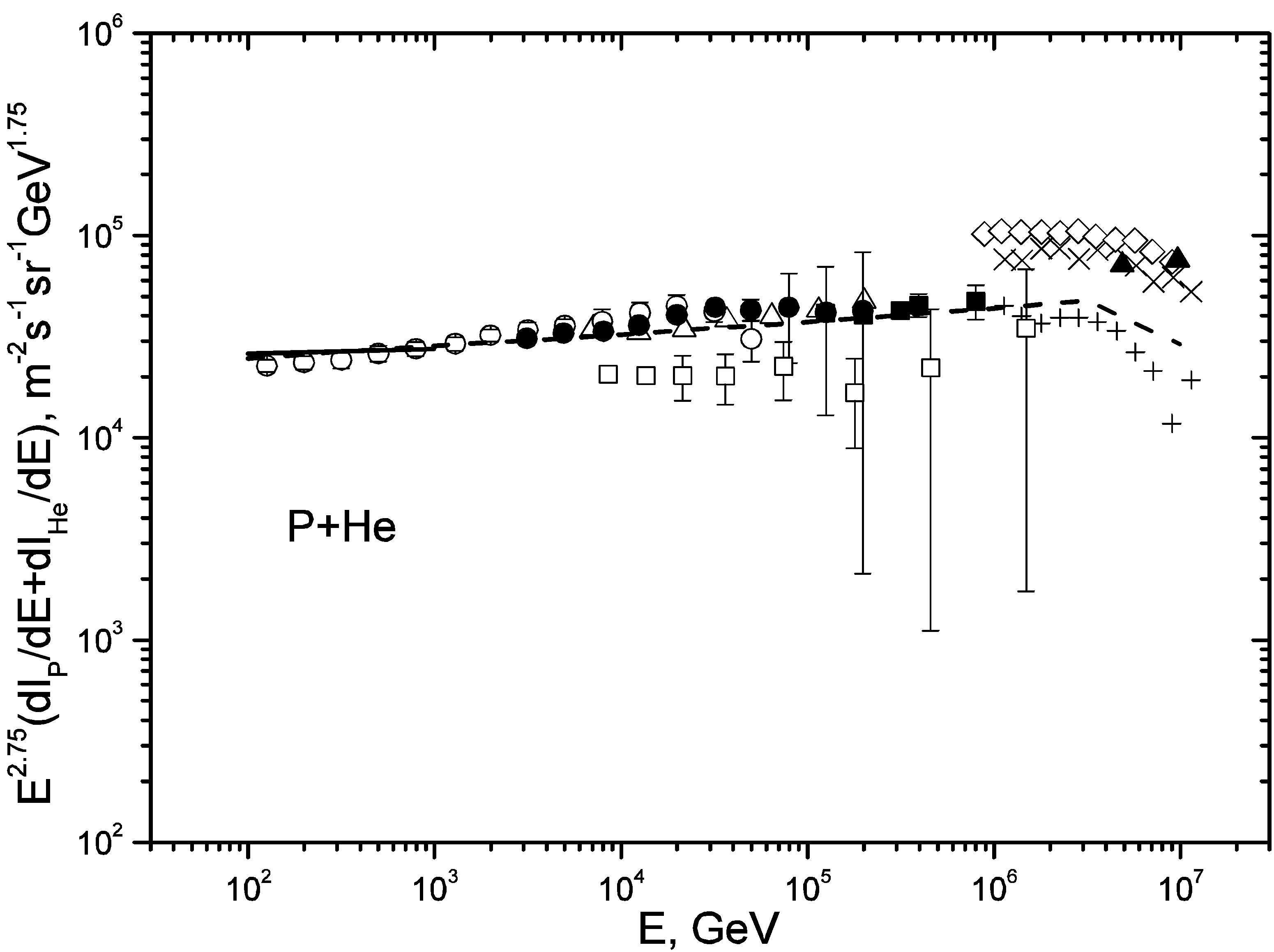}
\caption{The primary spectrum of protons and helium nuclei (p+He). Dashed line - modified Gaisser-Honda~[13] approximation. Data: solid line - AMS02~[14]; $\circ$~ATIC~2~[15]; $\bullet$~CREAM~[16]; $\blacksquare $~WCFTA~[17]; $\vartriangle $~ARGO~[18]; $\square $~RUN~JOB~[19]; $\diamond$~TUNKA~[21] (all particles); $\blacktriangle$~SPHERE~2~[22] (all particles); $\times$~KASKADE~[20] (all particles, QGSJET~II-03); $+$~KASKADE~[20] (all particles, SIBYLL~2.1).}
\label{fig-1}       
\end{figure}
\begin{figure}
\centering
\sidecaption
\includegraphics[width=8cm,height=6cm,clip]{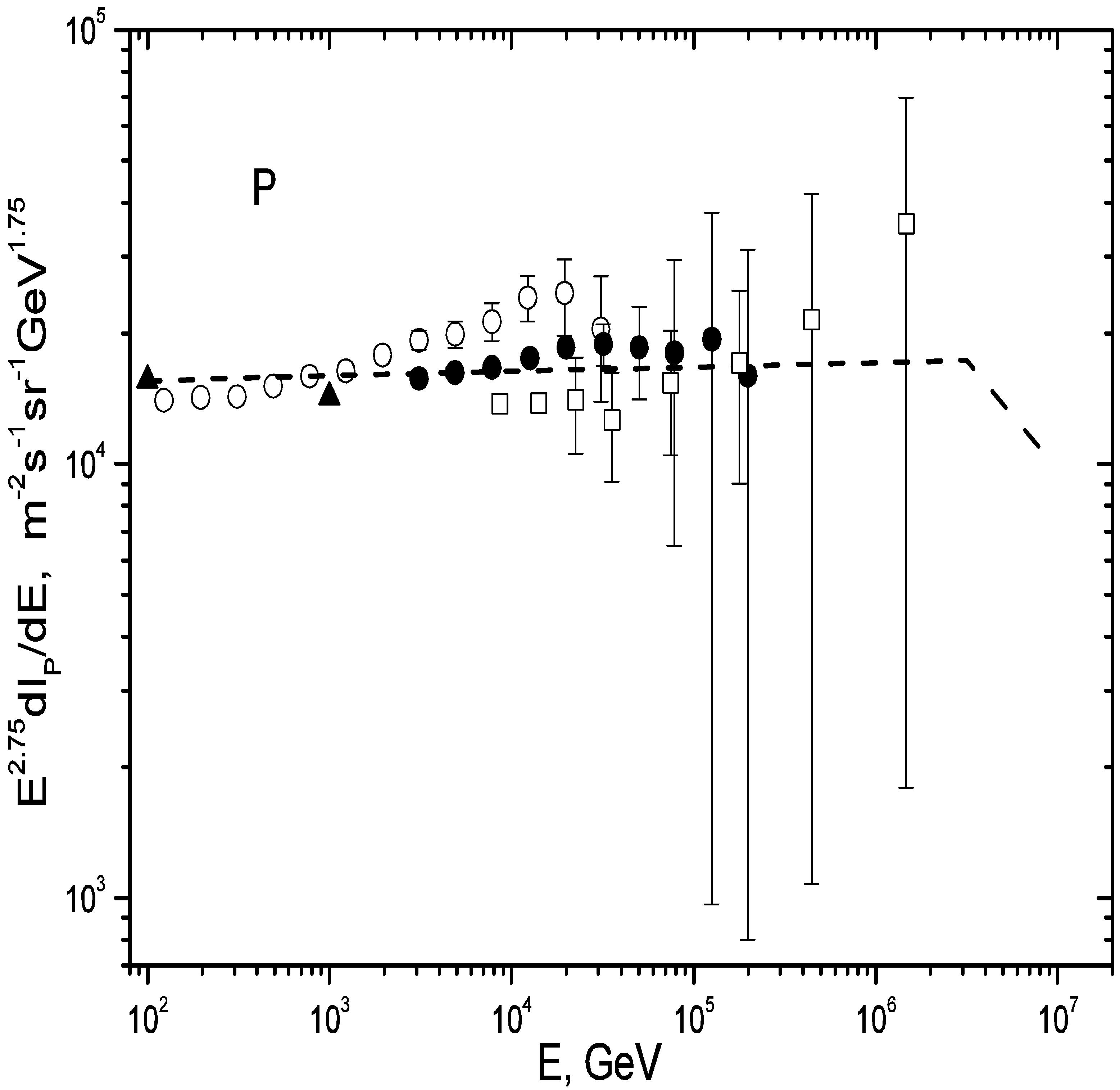}
\caption{The energy spectrum of the primary protons. Dashed line - modified Gaisser-Honda~[13] approximation. Various data - see the text. $\blacktriangle$~AMS02~[14]; $\circ$~ATIC~2~[15]; $\bullet$~CREAM~[16]; $\square $~RUN~JOB~[19].}
\label{fig-2}       
\end{figure}
\begin{figure}
\centering
\sidecaption
\includegraphics[width=8cm,height=6cm,clip]{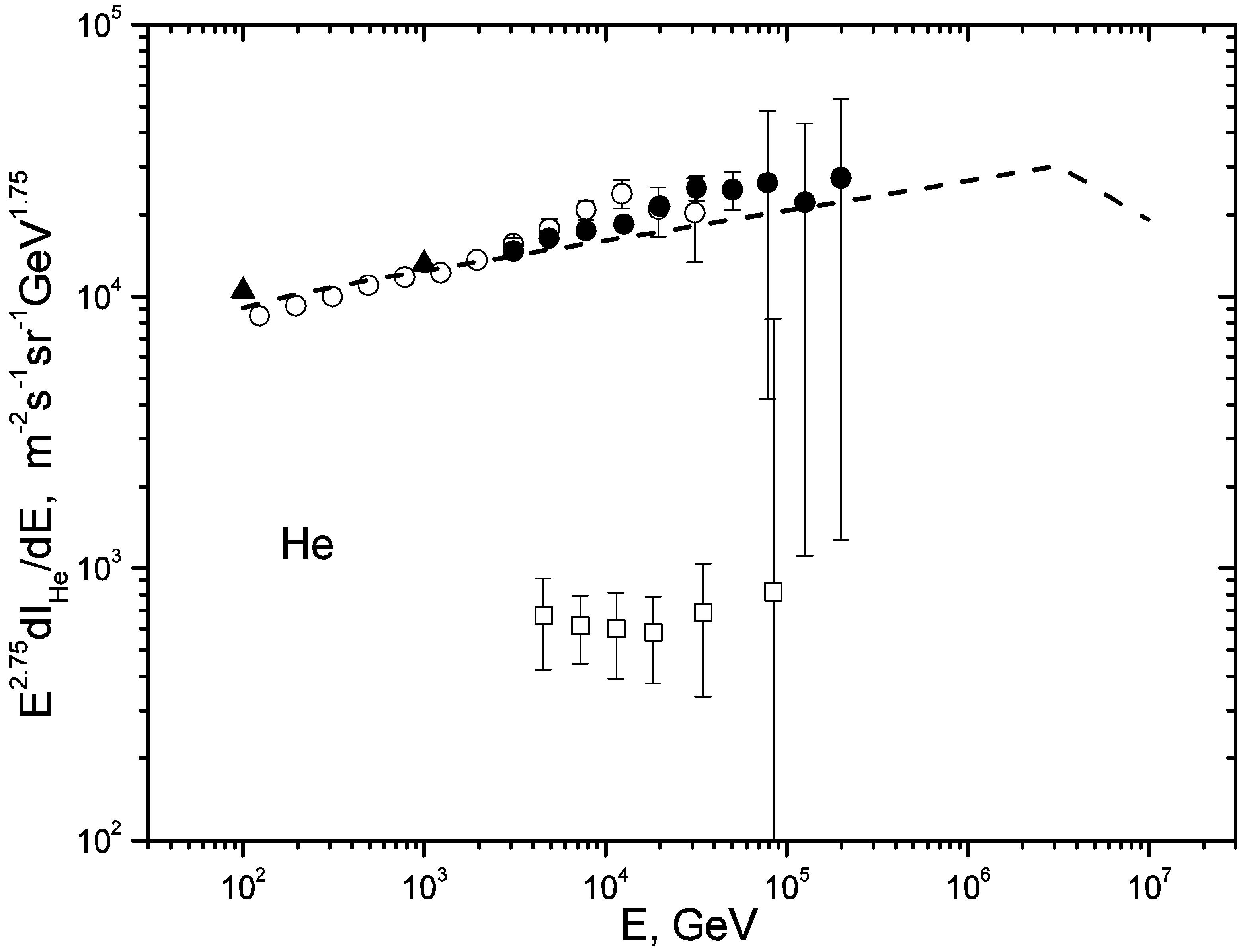}
\caption{The energy spectrum of the primary helium nuclei. Dashed line - modified Gaisser-Honda~[13] approximation. Various data - see the text. $\blacktriangle$~AMS02~[14]; $\circ$~ATIC~2~[15]; $\bullet$~CREAM~[16]; $\square $~RUN~JOB~[19].}
\label{fig-3}       
\end{figure}
\section{Method of calculation}
\label{sec-2}
The very simple original method of simulations of the energy spectrum of vertical muons can be described as follow [12]. 
Let the $(dI_p/dE)$ and $(dI_{He}/dE)$  be the differential energy spectra of the primary protons and helium nuclei. 
As for the spectrum of muons the energy per nucleon is of importance. 
So, the heavier nuclei give to this spectra negligible contribution. 
In the energy range of $10^2$ - $3\cdot 10^6$ GeV, we used the approximations $(dI_p/dE)_{GH}$ and $(dI_{He}/dE)_{GH}$ by Gaisser-Honda [13]. 
At energies above $E_{1}=3 \cdot 10^6 $ GeV, these spectra were multiplied by a factor $(E_1/E)^{0.5}$ and are refered as modified GH approximation. 
Figure 1 shows a comparison of the sum of modified GH approximations [13] for the primary protons and helium nuclei (dotted line) with the experimental data AMS02~[14] - solid line, ATIC2~[15] - hollow circles $(\circ)$, CREAM~[16] - dark circles $(\bullet)$, WCFTA~[17] - dark squares $(\blacksquare)$, ARGO~[18] - hollow triangles $(\vartriangle)$, RUN JOB~[19] - hollow squares $(\square)$. 
The experimental data KASKADE~[20], interpreted in terms of the model of QGSJET~II-03 shows the oblique crosses $(\times)$, and in terms of the model SIBYLL~2.1 - straight crosses $(+)$. 
The experimental data TUNKA~[21] for all primary particles are shown by hollow diamonds $(\diamond)$, and data SPHERE~2~[22] - dark triangles $(\blacktriangle)$. 
The spectra of the primary protons and primary helium nuclei are presented separately in Figure 2 and Figure 3 with various data mentioned above for a comparison. 
The notation: $(\blacktriangle)$  is applied in Figure 2 and Figure 3 only for the data of AMS02~[14]. 
From comparison with data it can be concluded that the accepted approximation [13] does not overestimate the flux of the primary protons and helium nuclei. 
This is important for the conclusions on possible uncertainties of tested models. 
The energy spectra of vertical muons $D_{p}(E_{\mu})dE_{\mu}$ and $D_{He}(E_{\mu})dE_{\mu} $ induced by the primary protons and helium nuclei are expressed by simple integrals over the energy $E$ of the primary particles as follows:
$$ D_{p}\left(E_{\mu}\right)\cdot dE_{\mu}=\int dE \cdot \left(\frac{dI_{p}}{dE}\right) \cdot S_{p}\left(E_{\mu},E\right) \cdot dE_{\mu}  $$
$$ D_{He}\left(E_{\mu}\right)\cdot dE_{\mu}=\int dE \cdot \left(\frac{dI_{He}}{dE}\right) \cdot S_{He}\left(E_{\mu},E\right) \cdot dE_{\mu}  $$
The sum of these spectra:
$$ D\left(E_{\mu}\right)=\left(D_{p}\left(E_{\mu}\right)+D_{He}\left(E_{\mu}\right)\right) $$ will be used for comparison with data [9-11].
Functions $S_{p}(E_{\mu},E) \cdot dE_{\mu}$ and  $S_{He}(E_{\mu},E) \cdot dE_{\mu}$ are the differential energy spectrum of muons in showers induced by the primary protons and helium nuclei with the fixed energy $E$. 
These spectra were calculated for 24 and 19 values of the energy $E $ of the primary protons and helium nuclei, respectively, in the range of $10^2 \div 10^7$ GeV. 
Calculations have been carried out in terms of two models of hadron interactions (QGSJET~II-03~[4] and QGSJET~II-04~[5]) using the package CORSIKA~7.4~[23]. 
The calculations were performed with the statistics of $10^6$ events in the low energy region and up to $10^2$ events at the highest energies of the primary particles. 
\begin{figure}
\centering
\sidecaption
\includegraphics[width=8cm,height=7cm,clip]{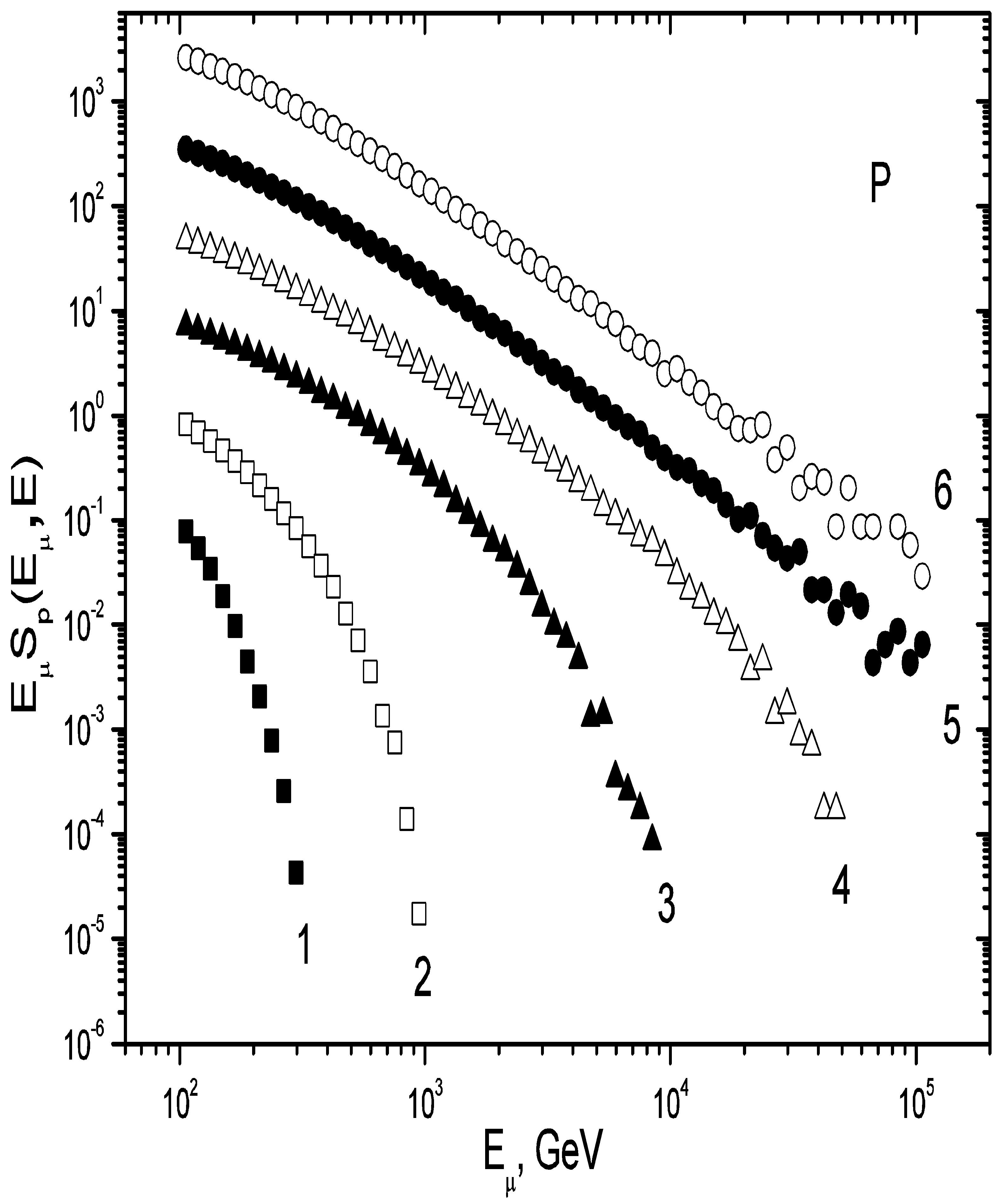} 
\caption{The energy spectra of muons in showers induced by the primary protons with various fixed energies $E $ (simulations in terms of QGSJET~II-04 model): $ 1-3,162 \cdot 10^2 \mbox{; } 2 - 10^3 \mbox{; } 3 - 10^4 \mbox{; } 4 - 10^5 \mbox{; } 5 - 10^6 \mbox{; } 6-10^7$ GeV.}
\label{fig-4}       
\end{figure}
\begin{figure}
\centering
\sidecaption
\includegraphics[width=8cm,height=7cm,clip]{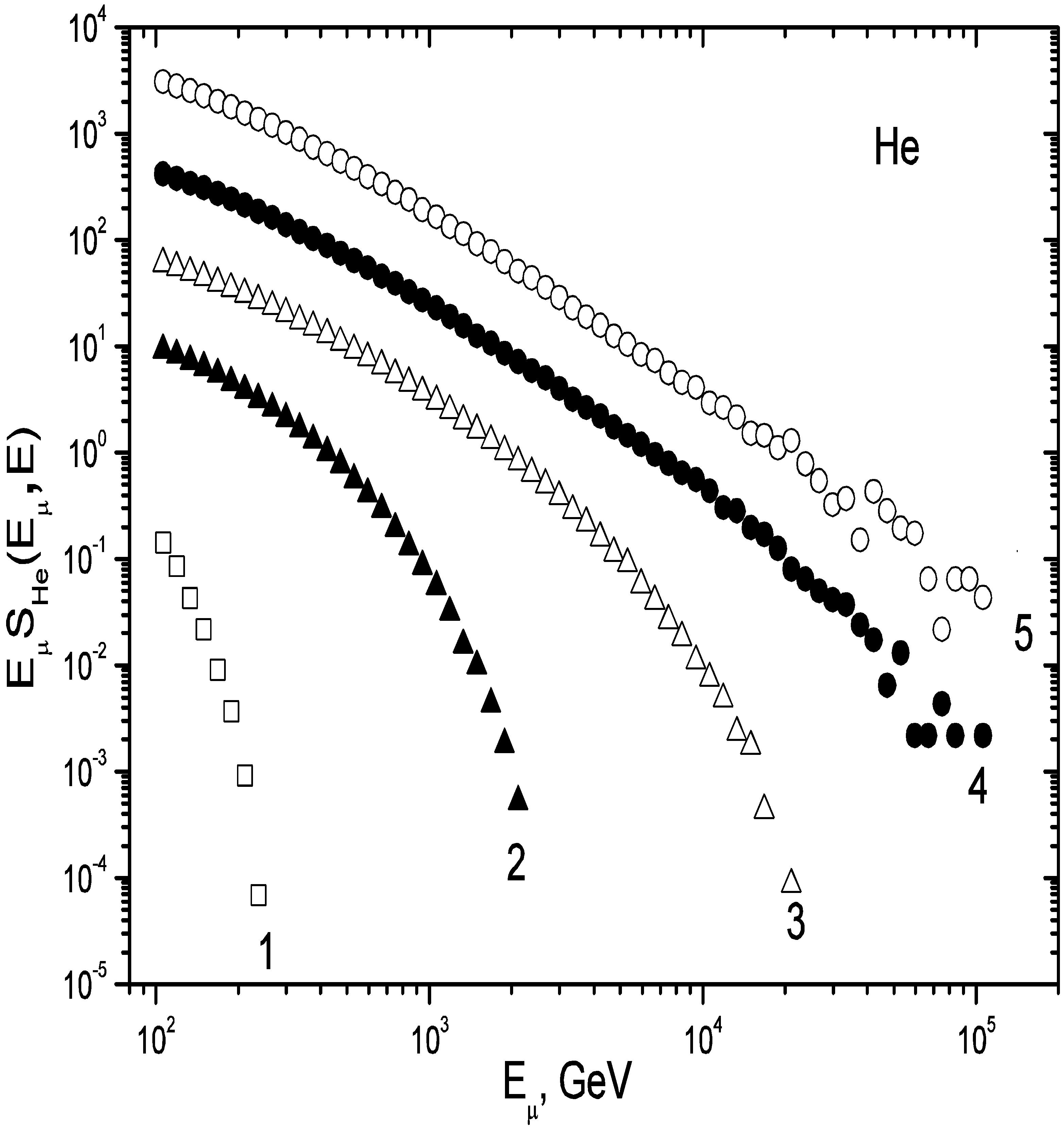}
\caption{The energy spectra of muons in showers induced by the primary helium nuclei with various fixed energies $ E $ (simulations in terms of QGSJET~II-04 model): $ 1 - 10^3 \mbox{; } 2 - 10^4 \mbox{; } 3 - 10^5 \mbox{; } 4 - 10^6 \mbox{; } 5 - 10^7 $ GeV.}
\label{fig-5}       
\end{figure}
Figure 4 and Figure 5 show examples of the muon energy spectra  $S_{p}\left(E_{\mu},E\right)$ and $S_{He}\left(E_{\mu},E\right)$ in the energy range of $10^2 \div 10^5$ GeV calculated in the terms of model QGSJET~II-04~[5] for the 6 values of proton energies and the 5 values of helium nuclei energies, respectively. 
It is seen that in the energy range of $10^4 \div 10^6$ GeV statistics is small. 
Therefore, we will use the energy range of $10^2 \div 10^4$ GeV for a comparison. 
Dependence of the spectra $S_{p}\left(E_{\mu},E\right)$ and $S_{He}\left(E_{\mu},E\right)$ on the models of hadron interactions (QGSJET~II-03 $(\bullet)$ and QGSJET~II-04 $(\circ)$)  are shown in Figure 6 and Figure 7 for the primary protons and helium nuclei, respectively for the energy $E=10^5$ GeV. 
It is evident that the model QGSJET~II-04~[5] predicts the greatest density of muons, while the model QGSJET~II-03~[4] - the lowest one. 

\section{The results of simulations}
\label{sec-3}
The spectra of vertical muons $D(E_{\mu})$ in the energy range of $10^2$ - $10^4$ GeV for the hadron interaction models QGSJET~II-03~[4] and QGSJET~II-04~[5] are presented in Figure 8. 
It can be seen that the model QGSJET~II-04 [5] predicts the intensity of muon flux by a factor $2 \div 3$ higher than the intensity of the muon flux calculated in terms of QGSJET~II-03~[4] model. 
This finding is consistent with the results shown in Figure 6 and Figure 7. 
Figure 8 is also very clearly demonstrates an increase in steepness of the muon spectrum at energies $E$ above 100 GeV. 
The constant $B_{\pi} \simeq 100$ GeV is the decay constant for $\pi$-mesons in the atmosphere. 
Therefore, the muon spectrum become steeper because $\pi$-mesons at energies $E > B_{\pi}$ rather interact with nuclei in the atmosphere than decay into muons. 
Comparison of the calculated spectra with experimental data allows us to test models. 
The ratios MC/DATA of the results of calculations for models [4] and [5] to the smooth approximation of the data of collaborations L3~+~Cosmic~[9], MACRO~[10] and LVD~[11] are shown in Figure 9. 
It can be seen that these ratios are increasing from $\sim $ 1.4 to $\sim $ 1.7 for the model QGSJET~II-04~[5] and are decreasing from $\sim $ 0.8  to $\sim $ 0.6 for the model QGSJET~II-03~[4] at muon energies in the range of $10^2 \div 10^4$ GeV. 
The most important fact is that these  increase becomes higher at energies $E_{\mu}$ above ${10}^3$ GeV for the QGSJET~II-04 model.
 For the QGSJET~II-03 model the ratios MC/DATA become constant at the level of $\sim 0.6$ at energies $E_{\mu}$ above $10^3$ GeV. 
No slowing of this increase is seen at higher energies of muons. 
Thus, Figure 9 demonstrates a very serious difference between the calculated spectra and the data reported in [9], [10] and [11]. 
This difference is associated with a different rate of energy fragmentation of projectile particles in events of its interactions with nuclei in the atmosphere.
Thus, the model QGSJET~II-03 underestimate the probability of secondary particles production at highest energies.
The model QGSJET~II-04 overestimate this probability of secondary particles production at the highest energies.
According to calculations, the main contribution to integrals $D_{p}\left(E_{\mu}\right)\cdot dE_{\mu}$ and $D_{He}\left(E_{\mu}\right)\cdot dE_{\mu}$ comes from secondary particles with energies in the ranges of (0.01 -- 0.6)$ E$ and (0.001 -- 0.1)$E$, where $E$ is the energy of a projectile particle for the primary protons and helium nuclei, respectively. 
The results of this comparison are also confirmed by the data of the LHCf~[6] and TOTEM~[7].
\begin{figure}
\centering
\sidecaption
\includegraphics[width=8cm,height=7cm,clip]{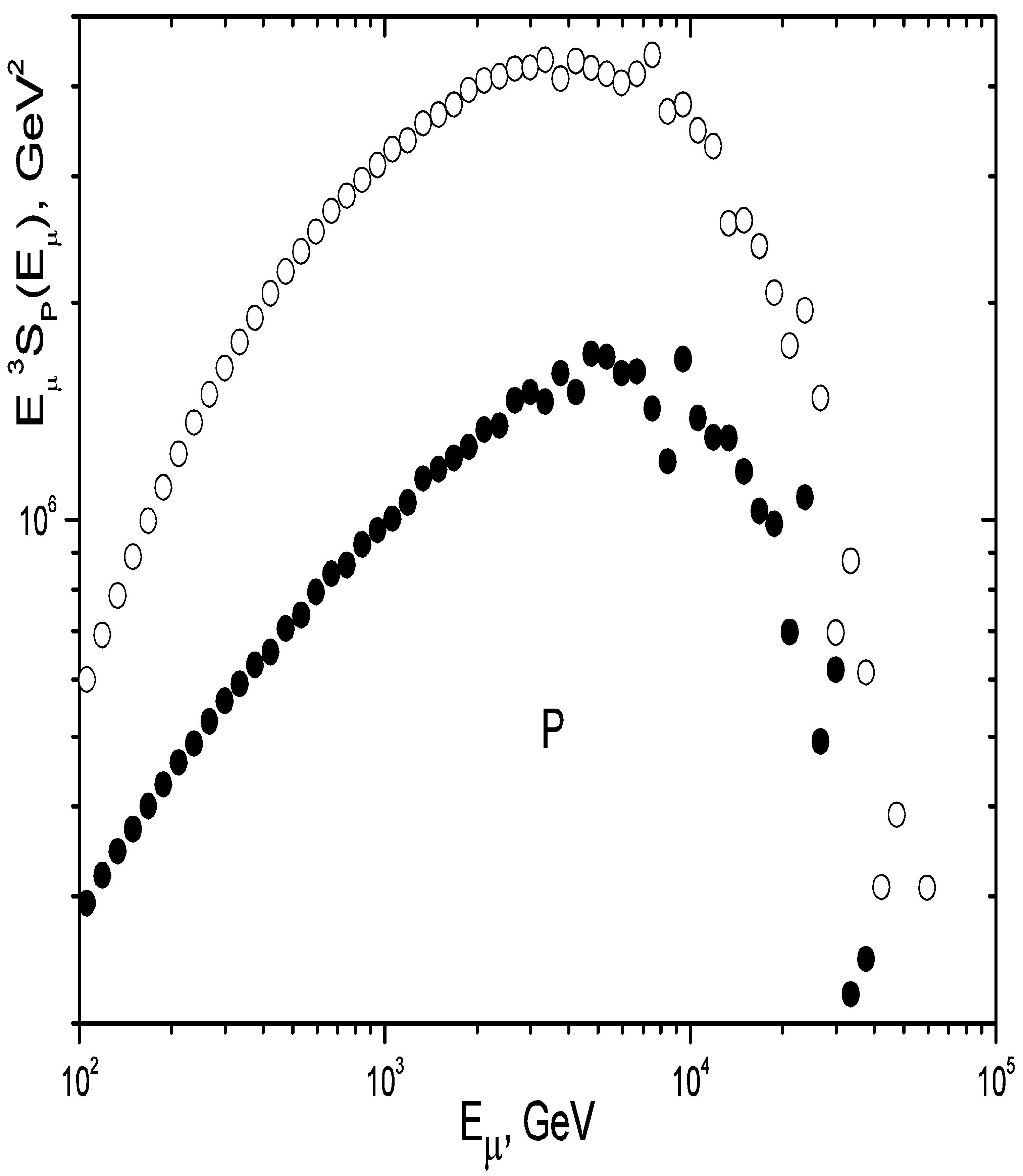}
\caption{The energy spectra of muons in showers induced by the primary protons with the fixed energy $E=10^5$ GeV calculated in terms of two models: $\bullet$ QGSJET~II-03, $\circ$ QGSJET~II-04.}
\label{fig-6}       
\end{figure}
\begin{figure}
\centering
\sidecaption
\includegraphics[width=8cm,height=7cm,clip]{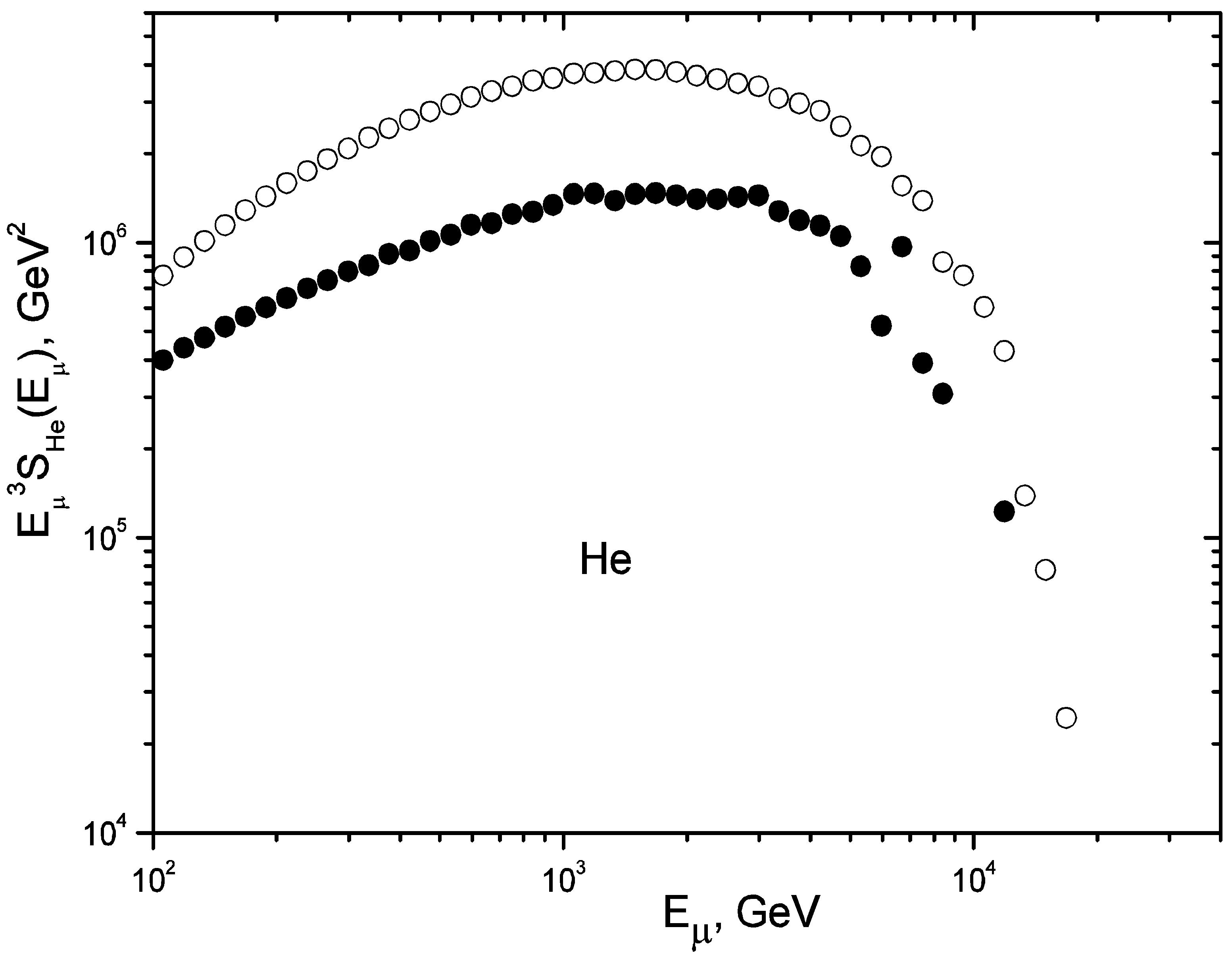}
\caption{The energy spectra of muons in showers induced by the primary helium nuclei with the fixed energy $E=10^5$ GeV calculated in terms of two models: $\bullet$ QGSJET~II-03, $\circ$ QGSJET~II-04.}
\label{fig-7}       
\end{figure}
For example, the QGSJET~II-04 model~[5] overestimates the density of charged particles $dN_{ch}/d \eta$ per unit of pseudorapidity at the pseudorapidity $\eta = 6.345$ by a factor of $k \approx 1.3$ as compared to the TOTEM data [7].
This difference increases at large values $\eta$ because of the difference between the slopes of the calculated curve and the data from [7]. 
The QGSJET~II-04~[5] model predicts the density $dN_{ch}/d{\eta}$ which is (18--30)$\%$ higher than the data [7] in the range $5.3 \le \eta \le 6.4$. 
It is also important to note that for difference of rapidity $\Delta y \simeq 0$, the value of the average transverse momentum $\left<p_{T}\right>$ for $\pi^{0} $-mesons is about 50 MeV/c less than data [24]. 
Under the assumption that a similar decreasing of the $\left<p_{T}\right>$ dependence is also valid for charged $\pi$ mesons. 
The calculated density of muons at large distances from the axis shower will be underestimated. 
This underestimation was observed by the Pierre Auger Collaboration [25] and in Yakutsk [26]. 
The energy spectra of photons in $p$--$p$ collisions at the energy of $\sqrt{s}= 7$ TeV in the pseudorapidity range $8.81 \le \eta \le 8.99$ are $2\div4$ times above the one's predicted by the QGSJET~II-03 model [4]. 
So, all these models should be significantly corrected at the highest energies of secondary particles.  
\begin{figure}
\centering
\sidecaption
\includegraphics[width=8cm,height=6cm,clip]{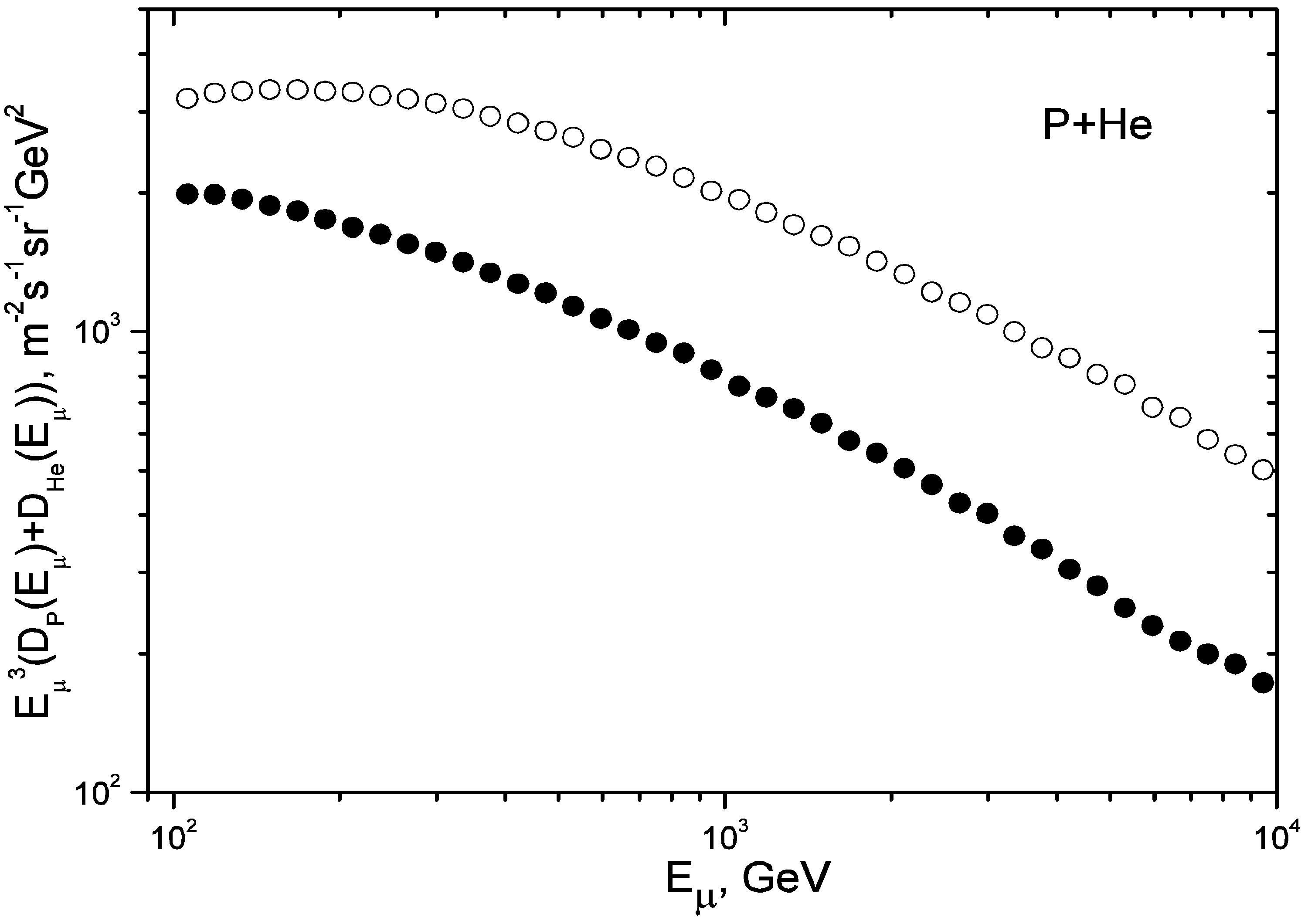}
\caption{The calculated energy spectra of near vertical muons for various models: $\bullet$ QGSJET~II-03, $\circ$ QGSJET~II-04.}
\label{fig-8}       
\end{figure}
\begin{figure}
\centering
\sidecaption
\includegraphics[width=8cm,height=6cm,clip]{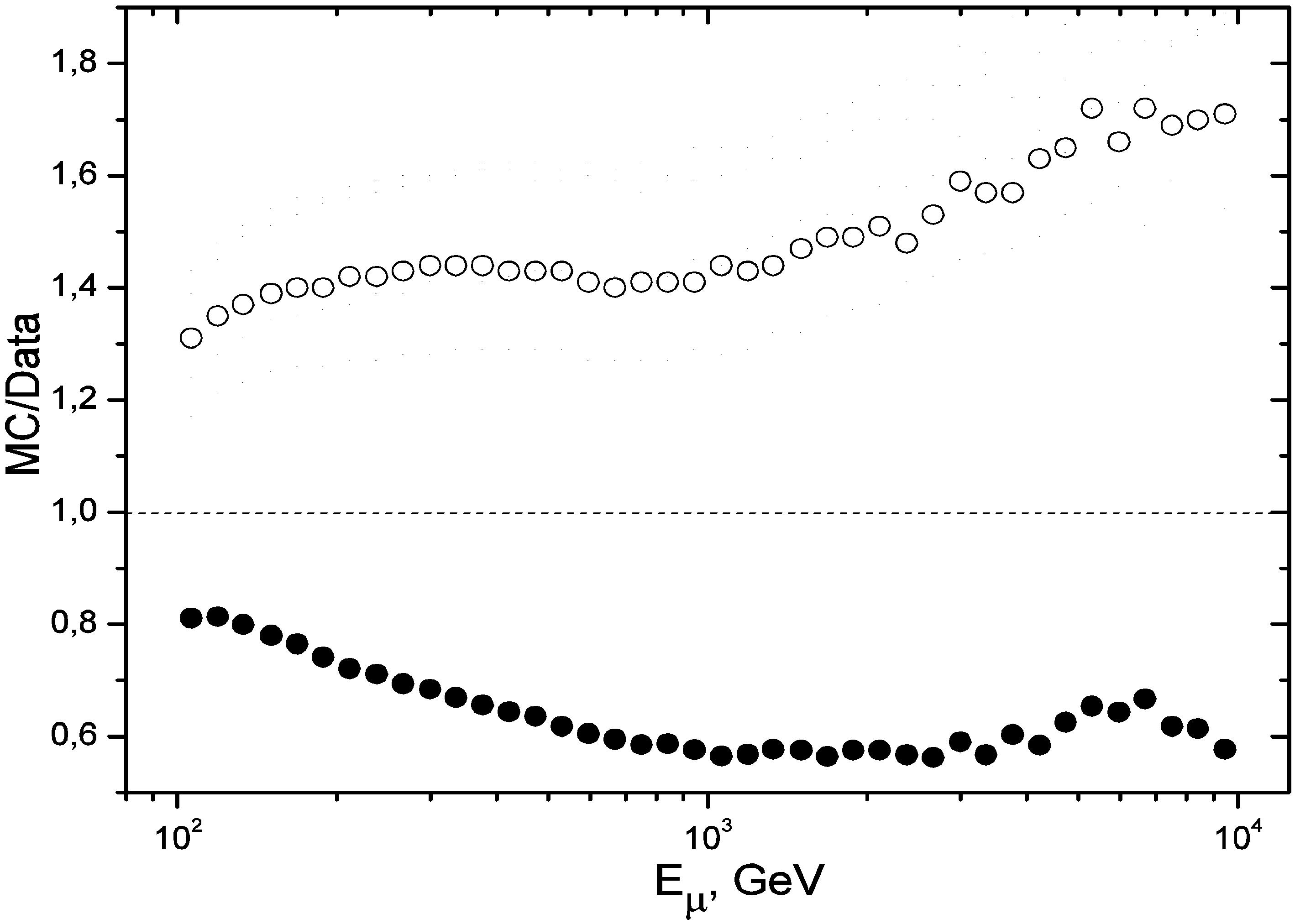}
\caption{Comparison of calculated muon energy spectrum with data~[9-11]. The ratios MC/DATA are shown: $\bullet$ QGSJET~II-03, $\circ$ QGSJET~II-04.}
\label{fig-9}       
\end{figure}
\section{Conclusion}
\label{sec-4} 
The atmospheric muon energy spectra calculated in terms of the QGSJET~II-04~[5] model is by a factor $f=1,7$ higher then data [9], [10], [11] at energy $E_{\mu}=10^4$ GeV.  The atmospheric muon energy spectrum calculated in terms of the QGSJET~II-03~[4] is by a factor $f=1,5$ lower then data [9], [10], [11] at energy $E_{\mu}=10^4$ GeV. So, we can conclude, these models of hadronic interactions should be updated at very high energies of secondary particles. 

\begin{acknowledgement}{}
\label{acknowledgements}
Authors thank LSS grant (grant 3110.2014.2) for support! Speaker thank organizing committee of the ISVHECRI 2014 for well organised conference and administrative support! 

\end{acknowledgement}

\end{document}